\documentclass[aps,prl,reprint,groupedaddress,showpacs,amsmath,amssymb]{revtex4-1}
\usepackage{epsfig}
\usepackage{dcolumn}
\usepackage{bm}

\newcommand{\degs}{$^{\circ}$}
\newcommand{\Ul}{$^{\mathrm{238}}$U$^{\mathrm{4+}}$}
\newcommand{\Uh}{$^{\mathrm{238}}$U$^{\mathrm{28+}}$}

\begin{document}


\title{Experimental Proof of Adjustable Single-Knob Ion Beam Emittance Partitioning}


\author{L.~Groening, M.~Maier, C.~Xiao, L.~Dahl, P.~Gerhard, O.K.~Kester, S.~Mickat, H.~Vormann, M.~Vossberg}
\affiliation{GSI Helmholtzzentrum f\"ur Schwerionenforschung GmbH, Darmstadt D-64291, Germany}
\email[]{la.groening@gsi.de}
\author{M.~Chung}
\affiliation{Ulsan National Institute of Science and Technology, Ulsan 698-798, Republic of Korea}


\date{\today}

\begin{abstract}
The performance of accelerators profits from phase space tailoring by coupling of planes. The previously applied techniques swap the emittances among the three planes but the set of available emittances is fixed. In contrast to these emittance exchange scenarios the emittance transfer scenario presented here allows for arbitrarily changing the set of emittances as long as the product of the emittances is preserved. This letter is on the first experimental demonstration of transverse emittance transfer along an ion beam line. The amount of transfer is chosen by setting just one single magnetic field value. The envelope-functions (beta) and -slopes (alpha) of the finally uncorrelated and re-partitioned beam  at the exit of the transfer line do not depend on the amount of transfer.
\end{abstract}

\pacs{41.75.Ak, 41.85.Ct, 41.85.Ja, 41.85.Lc}

\maketitle


\section{Introduction}
\begin{figure}[hbt]
\centering
\includegraphics*[width=86mm,clip=]{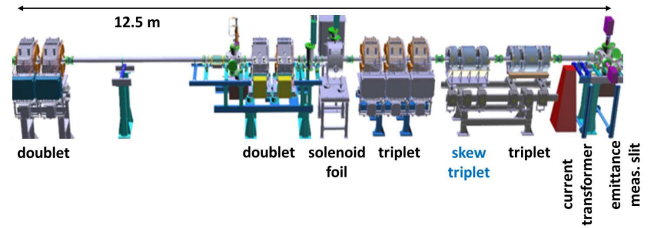}
\caption{(color online) Beam line of EMTEX (Emittance Transfer Experiment).}
\label{beamline}
\end{figure}

\begin{figure}[hbt]
\centering
\includegraphics*[width=86mm,clip=]{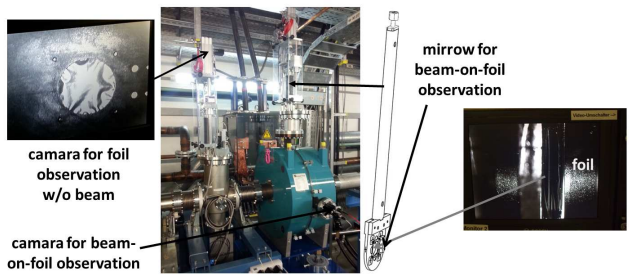}
\caption{(color online) The EMTEX solenoid with an actuator to place the stripping foil in its magnetic center. The foil can be observed in-situ by a camera from the right w.r.t. beam direction. Additionally, it can be inspected from the rear when the beam is off.}
\label{solenoid}
\end{figure}

Important figures of merit for any accelerator are the beam emittances it provides at its exit. Emittances are measures for the amount of phase space being occupied by the particle distribution. Along an ideal linear focusing lattice the dynamics in the three planes are not coupled and the emittances remain constant in each plane. Unavoidable coupling from fringe fields or from dispersion increases the emittances in the coupled planes. Accordingly, lattice elements were designed in order to minimize the coupling, except for dedicated applications such as spectrometers for instance. However, the case may rise that the performance of an accelerator is well within the emittance budget in one plane but far beyond in another one. Exchange of emittances between two planes may be sufficient to remain within all emittance budgets. Such schemes must involve inter-plane coupling.\\
Exchange of the two transverse emittances by a quadrupole triplet rotated by 45\degs (skew triplet) is state-of-the-art. Edwards et al.~\cite{Edwards} proposed creation of beams with strongly different transverse emittances, i.e. round-to-flat adaptors, and references~\cite{Brinkmann,Piot_prstab2006} report on successful demonstration. They used the solenoidal field of the electron source as inter-plane coupling element as well as a subsequent skew triplet. Application of this concept to ion beams is proposed in~\cite{Bertrand,Groening_prstab,Xiao_nim}. Exchange between the longitudinal and one transverse plane was considered and/or demonstrated in~\cite{Emma_prstab2006,Sun_prl2010,Chao_prstab2011}. Coupling was provided through installation of a transversely deflecting cavity inside a dispersive section created by bending magnets. Several studies were done aiming at finding the best suiting coupling scenarios to optimize the emittance exchange~\cite{carlsten_prstab2011may} or to combine it with strong bunch length compression~\cite{carlsten_prstab2011aug}. A recent review of longitudinal to transverse emittance exchange beam lines can be found in~\cite{Thangaraj}.\\
A common property of all emittance shaping beam lines operated so far is that the set of values of the final emittances cannot be changed. Exchange beam lines just re-assign the emittance values to the planes. Final emittances of round-to-flat adapters are given by the field strength of the solenoid at the source. This field strength is an intrinsic part of the particle source design and must not be changed during operation. The set of final emittances achieved by these beam lines is called the set of eigen-emittances~\cite{Dragt}. The product of the $n$ eigen-emittances is equal to the $2n$ dimensional rms emittance of the beam. The beam lines applied so far preserve the set of eigen-emittances.\\
Transverse emittance transfer as proposed in~\cite{Groening_prstab,Xiao_prstab} instead offers a tool to arbitrarily choose the final emittances as long as their product is kept constant. It changes the set of eigen-emittances. The beam line is sketched in Fig.~\ref{beamline}.
It comprises as key components a solenoid with a stripping foil in its center and a skew triplet. Change of charge state in the solenoid changes the set of eigen-emittances and the skew quadrupoles serve to remove inter-plane correlations. The solenoid field strength determines the amount of transfer between the eigen-emittances. A detailed description of the concept and the beam line itself is given in~\cite{Groening_prstab, Xiao_prstab, Xiao_HB2012}.
It must be stressed that charge state stripping is state-of-the-art in ion linacs. For intermediate to heavy mass ions it is even mandatory in order to perform acceleration at reasonable efficiency.
The so-called EMTEX set-up (Emittance Transfer Experiment) has two very convenient features. The optics of the beam line behind the solenoid does not depend on the solenoid field strength. It provides excellent decoupling if it is set for a solenoid field $B_0$ and if the applied field $B$ satisfies $|B|\leq |B_0|$. Additionally, the Twiss parameters $\beta$ and $\alpha$ provided at its exit do not change with the field strength $B$. These very comfortable decoupling and re-matching features were observed in simulations~\cite{Xiao_prstab} and were analytically explained later in~\cite{Groening_arxiv}. Hence EMTEX is an one-knob tool for transverse emittance partitioning. The knob is the solenoid field strength. As shown in Fig.~\ref{beamline} EMTEX comprises two doublets to match the beam to the entrance of the solenoid. The foil can be moved into the solenoid center and it can be observed by two camaras as shown in Fig.~\ref{solenoid}. The solenoid provides inter-plane coupling.
After the solenoid the coupled beam is matched by a triplet to the skew triplet that provides decoupling. Another regular triplet re-matches the beam for further transport. It is followed by a slit-grid beam emittance meter which allows for measuring the phase space distribution in the horizontal and vertical planes.\\
In the next section the experimental procedure will be described, followed by presentation of results from measurements and of comparison with predictions from theory and with tracking simulations. The paper closes with a summary and an outlook.

\section{Experimental Procedure}
EMTEX was installed~\cite{Maier_IPAC2014} last year along the transfer channel from GSI's UNIversal Linear ACcelerator (UNILAC)~\cite{UNILAC} to the synchrotron.
For the experiment a low intensity beam of $^{\mathrm{14}}$N$^{\mathrm{3+}}$ at 11.4~MeV/u was used. The relative momentum spread of the beam was less than 10$^{\mathrm{-3}}$. First, all EMTEX magnets were turned off and the stripping foil was removed from the solenoid. Full beam transmission was assured by using beam current transformes installed in front of and behind EMTEX. In order to assure that the beam in front of EMTEX does not exhibit already some inter-plane correlations, its image was observed using a fluorescence screen. The image was observed under variation of a quadrupole being installed in front of the screen. The quadrupole strength was varied to deliver a large spectrum of beam spot aspect ratios. Considerable inter-plane correlations would have manifested through tilted beam images. As only upright images were observed, it was assumed that no inter-plane correlations are present at the entrance to EMTEX.\\
In the following step, beam emittances were measured in both planes at the exit of EMTEX. The obtained Twiss parameters together with the settings of the first two doublets were used to determine the beam Twiss parameters at the entrance to the first doublet of EMTEX. A horizontal (vertical) rms emittance of 1.04 (0.82)~mm~mrad was measured.
Both doublets were set to provide a small beam with double waist at the location of the foil. The foil (carbon, 200~$\mathrm{\mu g}$/cm$^{\mathrm{2}}$, 30~mm in diameter) was moved into the solenoid. The measured beam current transmission increased by a factor of 2.3 as expected from charge state stripping of the beam from 3+ to 7+. Beam phase probes behind EMTEX were used to detect eventual beam energy loss induced by the foil. Within the resolution of the probes we assume that energy loss is close to the calculated value of 0.026~MeV/u from the ATIMA code~\cite{atima}. The same code was used to calculate the mean angular scattering of 0.474~mrad per plane. After determining all required input parameters the solenoid field was set to 0.9~T. Applying a numerical routine the three triplets behind the solenoid were set to decouple the beam and to provide for a beam with small vertical and large horizontal emittance together with full transmission through the entire set-up. These gradients were set and full beam transmission was preserved. Just very few steering was needed to re-center the beam in the emittance meter. This was required due to slight misalignment of the solenoid axis w.r.t. the beam axis. For the setting mentioned above both transverse emittances were measured. Afterwards the solenoid field was reduced stepwise to 0~T. 
The solenoid field $B_{i}$ was set by following the remanence-mitigating path
$B_{i-1}\rightarrow B_{max}\rightarrow 0~T\rightarrow B_{i}$.
All quadrupole gradients were kept constant. For each solenoid setting full transmission was preserved and both emittances were measured.

\section{Results}
\begin{figure}[hbt]
\centering
\includegraphics*[width=86mm,clip=]{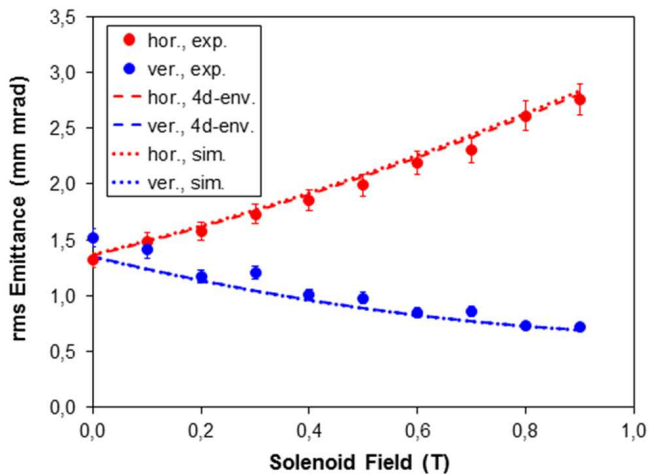}
\caption{(color online) Vertical (blue) and horizontal (red) rms emittances at the exit of the EMTEX beam line as functions of the solenoid field strength. All other settings were kept constant. Shown are results from measurements (dots), from application of the 4d-envelope model for coupled lattices (dashed), and from tracking simulations (dotted).}
\label{em_v_bsol_pmp}
\end{figure}

\begin{figure}[b]
\centering
\includegraphics*[width=86mm,clip=]{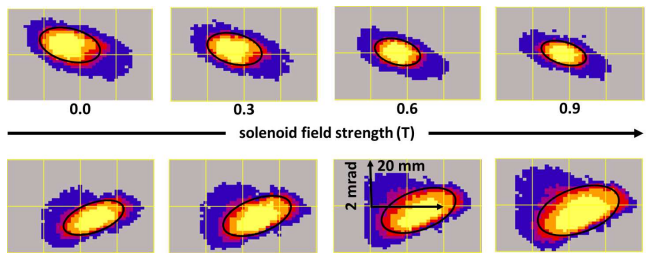}
\caption{(color online) Vertical (upper) and horizontal (lower) phase space distributions measured at the exit of the EMTEX beam line as functions of the solenoid field strength. All other settings were kept constant. Black ellipses indicate the 4$\times$rms ellipses.}
\label{phsp}
\end{figure}

\begin{figure}[h]
\centering
\includegraphics*[width=86mm,clip=]{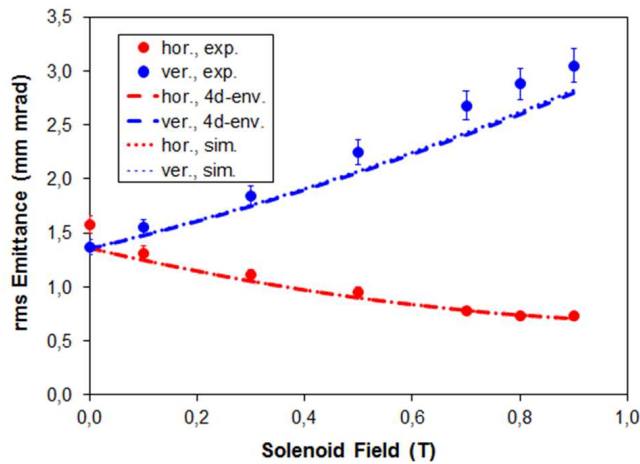}
\caption{(color online) Vertical (blue) and horizontal (red) rms emittances at the exit of the EMTEX beam line as functions of the solenoid field strength. All other settings were kept constant. Shown are results from measurements (dots), from application of the 4d-envelope model for coupled lattices (dashed), and from tracking simulations (dotted). W.r.t. Fig.~\ref{em_v_bsol_pmp} the gradients of the skew quadrupoles are inverted.}
\label{em_v_bsol_mpm}
\end{figure}

Figure~\ref{em_v_bsol_pmp} plots the measured rms emittances behind EMTEX as functions of the solenoid field strength. 
With increasing solenoid field the vertical emittance decreases while the horizontal increases. The product of the two emittances remains constant within the precision of the measurement. This behaviour is in full agreement to theoretical predictions from~\cite{Xiao_prstab} and to tracking simulations with TRACK~\cite{TRACK} using magnetic field maps. It is also in agreement with calculations that apply the recently developed 4d-envelope model for coupled lattices~\cite{Qin_prl2009,Qin_prl2013,Qin_prstab2014,Chung_up}. The observed emittance separation under variation of the solenoid field only, confirms that EMTEX is an one-knob tool for adjustable emittance partitioning.\\
Figure~\ref{phsp} displays measured phase space distributions as functions of the solenoid field strength.
It demonstrates that the shapes of the occupied areas in phase space and especially the shapes of the corresponding 4$\times$rms ellipses do not depend on the solenoid field strength within the resolution of the measurement. Also this experimental result is in full agreement to the observation from simulations reported in~\cite{Xiao_prstab} and to the properties of EMTEX derived analytically in~\cite{Groening_arxiv}. These references assumed a beam with exactly equal transverse emittances at the entrance of EMTEX. The beam emittances in the experiment differed by~23\% from each other. However, the quasi-invariance of the final ellipse shapes is in excellent agreement with the 4d-envelope model and with tracking simulations also for the present experiment. Accordingly, the experimental data also confirm that EMTEX is a one-knob emittance partitioning tool that preserves the beam envelope functions $\beta$ and $\alpha$ at its exit, if the initial beam emittances are similar. This feature makes it obsolete to re-match the envelopes as a function of the desired emittance partitioning once the partitioning is completed.\\
The emittance partitioning is given by the solenoid field strength. As shown in~\cite{Xiao_prstab}, inversion of the solenoid field swaps the behaviours of the emittances displayed in Fig.~\ref{em_v_bsol_pmp}, i.e. for negative solenoid field strengths the vertical emittance increases and the horizontal one decreases with the field strength. This could not be tested experimentally, since the solenoid power converter was uni-polar. However, inversion of the solenoid field strength is fully equivalent to inversion of the skew quadrupole gradients, while keeping all other gradients and the solenoid field constant. Inversion of the skew gradients corresponds to rotation of the skew quadrupoles by 90\degs , i.e. to swapping the transverse planes.
Accordingly, the sense of emittance partitioning is inverted for inverted skew gradients. This was verified experimentally as shown in Fig.~\ref{em_v_bsol_mpm}.
Also for inverted skew quadrupole gradients, preservation of the orientations and shapes of the measured phase space distributions was observed. Just the sizes of the corresponding 4$\times$rms ellipses changed with the solenoid field strength. For inverted skew quadrupoles the agreement to theory and to simulations is still good but slightly worse than for the case shown in Fig.~\ref{em_v_bsol_pmp}. Additionally, for a given solenoid field value the horizontal emittance values shown in Fig.~\ref{em_v_bsol_mpm} are not exactly equal to the vertical emittance values shown in Fig.~\ref{em_v_bsol_pmp}. According to theory they should be equal. However, the differences are very small. We attribute them to remanence effects in the solenoid and in the bi-polar skew triplets.

\section{Conclusion and Outlook}
Successful experimental demonstration of one-knob transverse emittance transfer was achieved. The predictions on the performance of the EMTEX set-up from analytical calculations and from tracking simulations with field maps were confirmed. Good to excellent agreement between calculations and measured data was observed. We believe that residual discrepancies are mainly from remanent field effects in the solenoid and in the bi-polar skew quadrupoles. Further reduction of remanence should be achieved by operating the magnets field-controlled instead of current-controlled. The latter was done in these experiments.\\
The presented experiment and theory do not include space charge. Additionally, we consider the stripping process to one single charge state. To clarify these points, simulations were done assuming an intense beam of \Ul at 1.4~MeV/u stripped to an average charge state of 28+ in a nitrogen gas jet. After dispersive charge state separation and filtering, the remaining \Uh ~beam of 15~emA pulse current causes a transverse tune depression of 40\% in the following drift tube linac (DTL). It turned out that the one-knob partitioning and matching features hold also under considerable space charge and even if the stripping process has a charge state spectrum. Further simulations indicated that such an intense flat beam can be accelerated to 11.4~MeV/u along a periodic DTL with reasonable preservation of beam quality~\cite{Groening_LINAC2014}.


\end{document}